# Optical control of cardiac cell excitability based on two-photon infrared absorption of AzoTAB

Shcherbakov D.[1,2,][*], Motovilov K.[2,3], Erofeev I.[2,3], Astafiev A.[1]

[1] N.N. Semenov Institute of Chemical Physics RAS, Moscow, Russia

[2] Department of General and Applied Physics, Moscow Institute of Physics and Technology, Dolgoprudny, Russia

[3] Research-Educational Center "Bionanophysics", Moscow Institute of Physics and Technology, Dolgoprudny, Russia

*Corresponding author: dimsney@gmail.com

## Abstract

Recent studies of AzoTAB (*2-{4-[(E)-(4 – etoxyphenyl ) diazinyl ]phenoxy} - N,N,N – threemethylammonium bromide* ) activity in excitable cell cultures have shown that this substance is able to control excitability depending on isomer, *cis* or *trans*, predominating in the cellular membrane. Control of isomerization can be performed noninvasively by UV-visual radiation. At the same time it is well-known that azobenezenes can be effectively transformed from one isomer into another by two-photon absorption. Current work is devoted to the study of *trans*-AzoTAB two-photon transformation in aqueous solution and inside primal neonatal contractive rat cardiomyocytes. In accordance with results obtained Azo-TAB can be used as a probe for two-photon optical control of cardiac excitability.

## Abbreviations

Azobenzene, AzoTAB, cardiomyocyte, two-photon absorption, excitability, photocontrol

**Introduction**

The derivatives of azobenzene are known as one of the best agents for optical control owing to very low photobleaching, high quantum yield and their ability to be photoisomerized repetitively. The difference between lengths of cis- and trans-isomers (4 and 4' positions) is as much as 4,4 Å [1-3]. This feature can be used for mechanical transfer of a functional group between two distinct positions depending on excitation radiation. On account of rather quick photoconversion between isomers occurring within picoseconds [4-6] the azobenzenes can be used in the studies devoted to the kinetics of macromolecules such as proteins. Also, azobenzene derivatives are widely used in different technological applications considering non-invasive remote control of surface properties, photo-patterning, nanofabrication, phto-actuation in liquid-crystalline polymers and thin films, micromechanics and microrobotics (please, see perfect review [7]).

First attempts to apply azobenzene photoswitches in biosystems were made in 1971 [8] when Erlanger and colleagues used azobenzene derivatives to control activity of acetylcholine receptor (nAChR). After this article azobenzene-based photoswitches were used for activity control of other ion channels including ionotropic glutamate receptor [9], voltage-gated potassium channel [10] etc. In some cases complex derivatives, containing reactive moiety of maleimide, which forms covalent bond with surface cysteines, were used.

AzoTAB was synthesized in 1994 by group of Japanese chemists in the course of their work devoted to the cationic derivatives of azobenzene [11]. Biochemical applications of azotab were studied much later. It was demonstrated that this cationic azo-surfactant allows photocontrol of bacteriorhodopsin conformational changes [12], reversible photocontrol of transcription/expression systems activity [13, 14] and may have other bio-applications.

At the same time in our lab it was shown that AzoTAB has unique combination of properties which make it perfect photoswitch for control of cardiac tissue excitation [15]. Coupling this method with optical mapping allows real-time 2D-modeling of different disorders of myocardium functioning [16].

AzoTAB use in living systems has two important limitations. The first one is toxicity of azobenezene fragment. It's almost unbreakable and can be overcome only through total change of photoswitch chemical structure. In our experiments we saw many times that cardiac tissue incubated with 100 µM AzoTAB loses its ability to counteract and produce active potential after 4-5 hours. And this effect doesn't depend on that which type of AzoTAB isomer, cis- or trans-, was presented in media most of the incubation time. The second limitation is absorption band of trans-AzoTAB with maximum around 350 nm. The depth of near UV penetration in biological systems is very low and doesn't exceed several dozens of microns [17]. There're several approaches which may help us to use AzoTAB in deeper layers of biological tissues. The first one is introduction of electron donors as substituents in azobenzene system. They would shift the absorption band of *trans*-AzoTAB into red region but simultaneously they would lower the activation energy of both back and direct transformations of azobenzene system. Usually, this leads to dramatic increase of the *cis*- to *trans*- transition rate and seems to be undesirable result. We can also partially stabilize *cis*-isomer by additional groups which form hydrogen bonds only in *cis*-state. This factor transforms molecule too much to keep the same affinity to the target.

The multiphoton absorption gives totally another key for solution of the problem, because it transfers the excitation band into infrared spectrum without adoption of the same old chemical structure which definitely works with the target. Azobenzenes show rather high molecular two-photon cross-section [18], but anyway it is several orders lower than normal absorption. Thus, the comparative stability of cis-isomer of AzoTAB (it takes several dozens of hours for cis-isomer to transform into trans-isomer in full darkness) makes two-photon approach probably usable for the case of this azobenzene derivative.

Current study was conducted to clarify next statements:
1. *Trans*-AzoTAB can be efficiently transformed into *cis*-AzoTAB by two-photon absorption both in solution and in living cardiomyocytes.
2. 100 µM AzoTAB toxicity to cells coupled with IR laser impact doesn't kill cells during common experiment time (1 hour).

3. The toxic effect of reactive oxygen species formed in water bulk phase under IR radiation in presence of AzoTAB can be sufficiently eliminated by ascorbic acid.

**Materials and methods**
1. *Excitable cardiac cell tissue preparation* was performed in accordance with method described in [15].
2. *Optical mapping of cardiac cells excitability* was made with calcium fluorescent dye fluo 4 [15, 16].
3. *AzoTAB was synthesized* in accordance with the method described in [11].
4. *The laser radiation of the cardiac cell culture* was performed using setup showed on the fig. 1. Petri dish was placed on a XY-scanning platform. AzoTAB was added to the solution in 100 µM concentration. Femtosecond laser beam was focused into the cells by an objective (NA=0.55). The cell culture surface area treated with laser was a square with 100X100 µm. The laser cross section had a round shape of 1.15 µm diameter. The time of full scan of the square by laser was 256 seconds. Thus, during the 30 min exposition period the laser made 7 cycles of cell culture irradiation. Observation of cell excitability restoration was conducted with the optical mapping setup based on Olympus IX71 microscope equipped with CCD-camera Andor iXon3 [16].

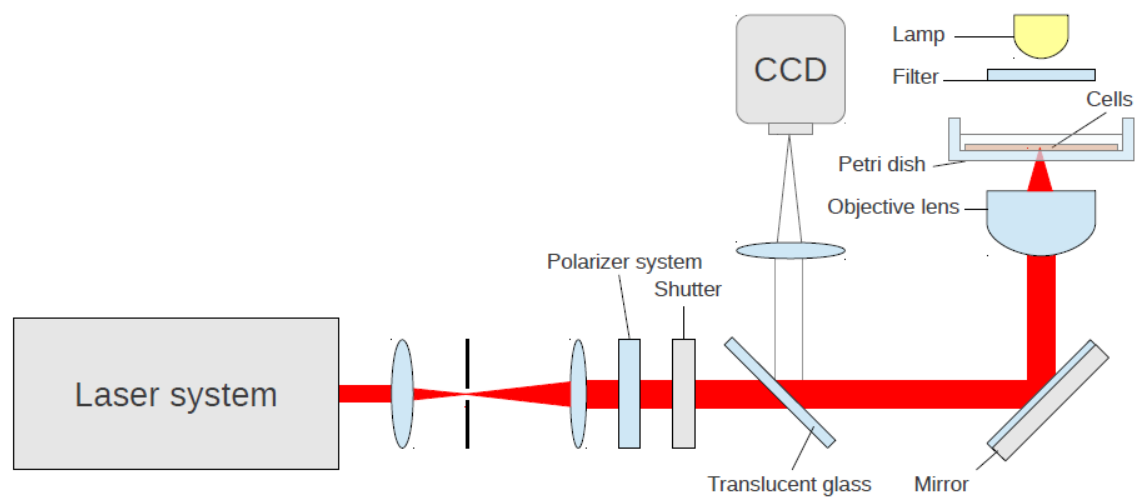

*Fig.1. Experimental set.*

5. A special experimental set for measuring two-photon absorption using z-scan method was assembled (fig. 2). The radiation source was a femtosecond Ti-sapphire laser (Tsunami, Spectra-Physics), pumped by a Nd:YAG laser (Millenia, Spectra-Physics) and generated pulses at 720 nm wavelength. The pulses passed through a regenerative amplifier (Spitfire) and a parametrical amplifier (Topas-white). The pulse duration after these amplifiers was 28 fs, with 1 Khz repetition rate and the energy up to 50 mkJ. The radiation was attenuated by a neutral filter for avoiding the effects related to supercontinuum generation. For more precise power adjustment a system of a rotary half-wavelength plate and a linear polarizer was used. Some of the radiation was reflected to a photodetector (Newport). The laser ray was focused by a convex lens with f=15 cm focal length at a 1 mm thick quartz cuvette with the sample substance. The energy of the pulses incident on the sample did not exceed 0,3 mkJ. The beam that passed the cuvette was collimated by the second lens and got to the second photodetector. Both photodetectors were connected to a power meter (Newport 2935-C). The digitized signal from both channels of the meter was registered and proceeded by a special PC program written in LabVIEW that calculated the averaged ratio of light powers on two detectors.

The cuvette with 3.8 mM AzoTAB solution in methanol was placed on a moving platform with a stepping motor that moved it along the beam with minimal step of 1 mkm. The platform was positioned by movement controller and the PC with the help of standard software. During the experiment the cuvette with the sample moved through the focus of the laser beam and for every coordinate the ratio of signals of photodetectors was measured.

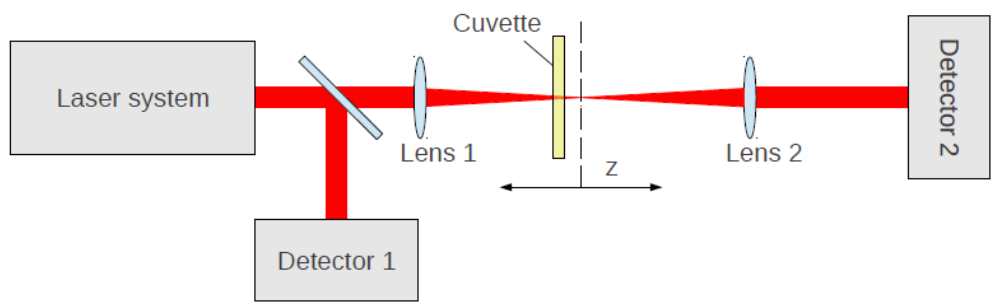

*Fig.2. Z-scan set.*

**Results and discussion**

Single photon absorption spectra of AzoTAB are depicted on fig.3. The trans- form does not absorb at the used laser wavelength (720 nm) but there is an absorption peak at double frequency (360 nm) that corresponds to two-photon absorption at the given wavelength.

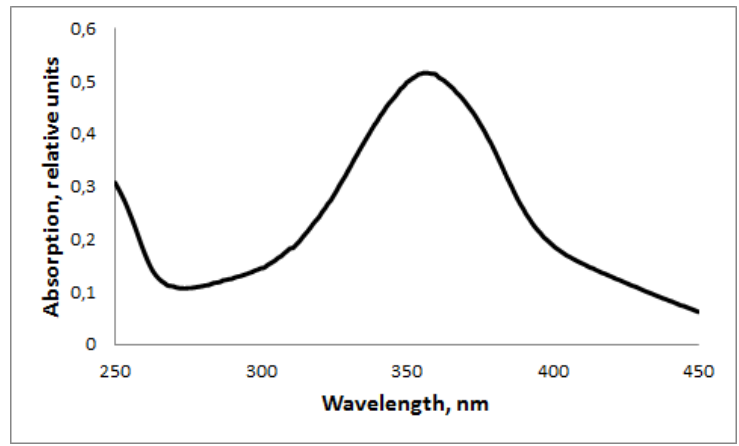

*Fig.3. AzoTAB one photon absorption spectrum*

According to [1], in case of low one photon absorption the normalized change in transmitted radiant energy ΔT(z) is related to sample coordinate by the equation:

$$\Delta T(z) \approx \frac{-\beta I_0 L}{2\sqrt{2}} \frac{1}{(1+z^2/z_0^2)} \quad (1)$$

where L is the cuvette thcikness, β is the two-photon absorption coefficient of the solution, $I_0$ is peak on-axis irradiance at the focus. The two-photon absorption cross-section can be found from β using the expression [1]:

$$\sigma_2 = \frac{hc\beta}{\lambda N} \quad (2)$$

The acquired z-coordinate dependence of the ratio of transmitted light to incident is given on fig.4. Here the absorption valley is clearly seen. According to expression (1) the absorption valley has Lorentzian profile. To define the focus depth and two-photon absorption coefficient we fit this dependence with Lorentzian function.

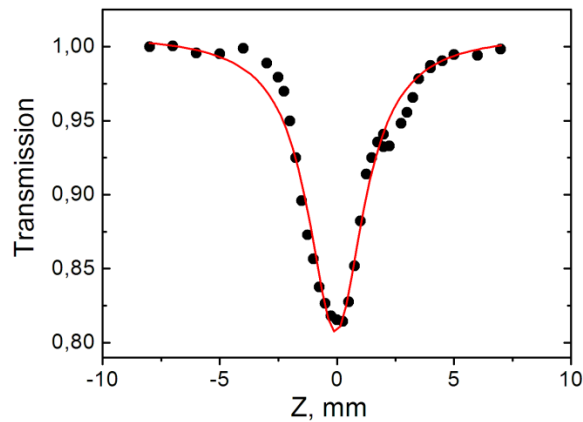

*Fig.4. Z-scan curve*

The defined nonlinear absorption coefficient and focus depth are:

$$\beta = 0{,}0194 \; cm/GW$$

$$z_0 = 1{,}44 mm$$

Using expression (2) we get the value of two-photon absorption cross-section $\sigma_2 = 23 GM$ for trans-AzoTAB.

**Estimation of stationary concentration during z-scan**

Though the studied solution showed two-photon absorption it is necessary to show that trans-isomer concentration was changed insignificantly during the measurement process. Given the two-photon absorption cross-section, we can calculate the rate of photochemical reaction occurring in the focal spot. Radiation power absorbed by the solution at a particular moment of time equals

$$\frac{\partial P}{\partial V} = \frac{2\sigma N}{h\nu} I^2$$

Let us now introduce the parameter A that determines the momentary photochemical reaction rate or the concentration change due to light absorption per time unit at a particular moment

$$A = \frac{\partial P}{\partial V}\Big|_{r=0} \cdot \frac{\gamma}{2h\nu N_A} = \frac{\sigma N}{(h\nu)^2 N_A} I^2(t) \ (5);$$

The intensity is nonzero only during the pulse and is zero nearly the whole time interval between the pulses, hence for time-average this equals

$$\bar{A} = \frac{\gamma \sigma c_{trans} I_0^2}{(h\nu)^2} \cdot \tau_0 \nu_{rep} = 66{,}8 (mol/(m^3 \cdot s))$$

If a new coefficient B is equal to

$$B = \frac{A}{c_{trans}} = \frac{\gamma \sigma I_0^2}{(h\nu)^2} \cdot \tau_0 \nu_{rep} = 1{,}74 (s^{-1})$$

then the dynamics if trans-form concentration change is described by following equation:

$$\dot{c}(\vec{r}, t) = D\Delta c(\vec{r}, t) - B(\vec{r}) c(\vec{r}, t)$$

$$\tau_x = \tau_y = \frac{d_x^2}{2D_{Ethanol}} = \frac{(18{,}2 \cdot 10^{-6})^2}{2 \cdot 5 \cdot 10^{-9}} = 3{,}3 \cdot 10^{-2} (s)$$

Now let us estimate the concentration change induced by two photon absorption:
$\Delta c \approx A\tau_x = 57{,}4 (mkmol/l)$

This value is negligible compared to the concentration in the cuvette. So the induced concentration change during z-scan process is not taken into account. Therefore, two photon cross-section of trans AzoTAB at 720 nm is 23 GM. Yet there is no experimental data on AzoTAB two-photon absorption rate, but as it was shown earlier [18], azoaromatic compounds may have various two photon cross sections ranging from $10^{-1}$ to $10^1$ GM.

Parameters of beam focusing depend not only on numerical aperture of the focusing objective but also on thickness of the laser beam incident on it. It is hard to measure the thickness of this beam, but it is known that it is nearly the same that the objective aperture. Hence the incident beam was considered as a planar wave with constant intensity. The Airy spot will be in the center of such beam. Yet, the beam is Gaussian, and for our calculations to proceed it is necessary to know its width at $1/e^2$ intensity. A Gaussian beam focus is generally approximated by an ellipsoid. We will consider it to be a prolate spheroid e.g. two of its main axes are equal ($d_x = d_y$). When Airy pattern is approximated with Gaussian the radius of Gaussian spot at $1/e^2$ intensity will be at a distance $w_0 = \frac{0{,}42\lambda}{NA}$ from the focus center. Here and further this value will be considered the beam waist radius. In our case (for NA=0,55):

$$w_0 = d_x = d_y = \frac{0{,}42\lambda}{NA} = 573 (nm)$$

Now let us find the focus depth $d_z$. It can be found through the relationship for non-paraxial Gaussian beams [19]:

$$\frac{d}{l} = \frac{1-\cos(\alpha)}{\sqrt{3-2\cos(\alpha)-\cos(2\alpha)}},$$

where d is focus width, and l is its depth. For NA=0,55 this ratio is 0,17, e.g. focus depth $d_z = 5{,}86 d_x = 3{,}4 (mkm)$.

## Calculating intensity at the beam focus center.

Beam power is an integral over coordinates of intensity in the focal plane:
$$P = \int_0^\infty I_0 \cdot 2\pi r e^{-\frac{2r^2}{w_0^2}} dr = \frac{\pi}{2} I_0 w_0^2,$$

Hence the intensity in the geometrical center of the focus is
$$\bar{I}_t = \frac{2P}{\pi w_0^2};$$

so the peak intensity is
$$I_0 = \frac{2P_l}{\pi w_0^2 \nu \tau} = 2{,}43 \cdot 10^{12} (W/cm^2)$$

## Diffusion coefficient estimation.

The diffusion coefficient of trans- form in water can be calculated with Stocks-Einstein formula. For the molecule of nm size we get
$$D = \frac{RT}{6\pi N_A \eta r} = 2{,}74 \cdot 10^{-5} \left(\frac{cm^2}{s}\right)$$

## Concentration distribution in the focal area.

The coefficient A should be calculated, substituting the obtained value of intensity into the expression (5) and taking the quantum yield to be 2.4:
$$A = \frac{\sigma c I_0^2}{(h\nu)^2} \cdot \tau_0 \nu_{rep.} = 7{,}8 \cdot 10^3 \left(\frac{mol}{l \cdot s}\right)$$

Now we obtain the coefficient B required for numerical calculation:
$$B = \frac{A}{c_{trans}} = 7{,}8 \cdot 10^7 \left(\frac{1}{s}\right)$$

In some time after the exposition started the trans- form concentration distribution in the focal spot will reach the stationary profile. These values can be found with the expression (6). The specific diffusion time over x and y axes can be calculated:
$$\tau_x = \frac{d_x^2}{2D} = \frac{(573 \cdot 10^{-9})^2}{2 \cdot 2{,}74 \cdot 10^{-9}} = 6{,}0 \cdot 10^{-5} (s)$$

Now, using (6), it is possible to estimate the concentration change in the center of the focus:
$$\Delta c_\infty \approx A\tau_x = 0{,}47 (mol/l)$$

This value is orders greater than the initial concentration. Hence the trans- isomer in the focal spot center will be depleted ant its concentration will be negligible. The radius of this low concentration zone can be estimated using the assumption that at its borders the two-photon induced concentration change is equal to initial concentration.
$$A_0 \tau_x e^{-2R^2/d_x^2} = c_0,$$

From where we get the value of low concentration zone radius R:
$$R = d_x \sqrt{\frac{1}{2} \ln \frac{A_0 \tau_x}{c_0}} = 2{,}4 (mkm)$$

We will illustrate it having calculated the concentration distribution numerically and considering it to be stationary on fig.5.

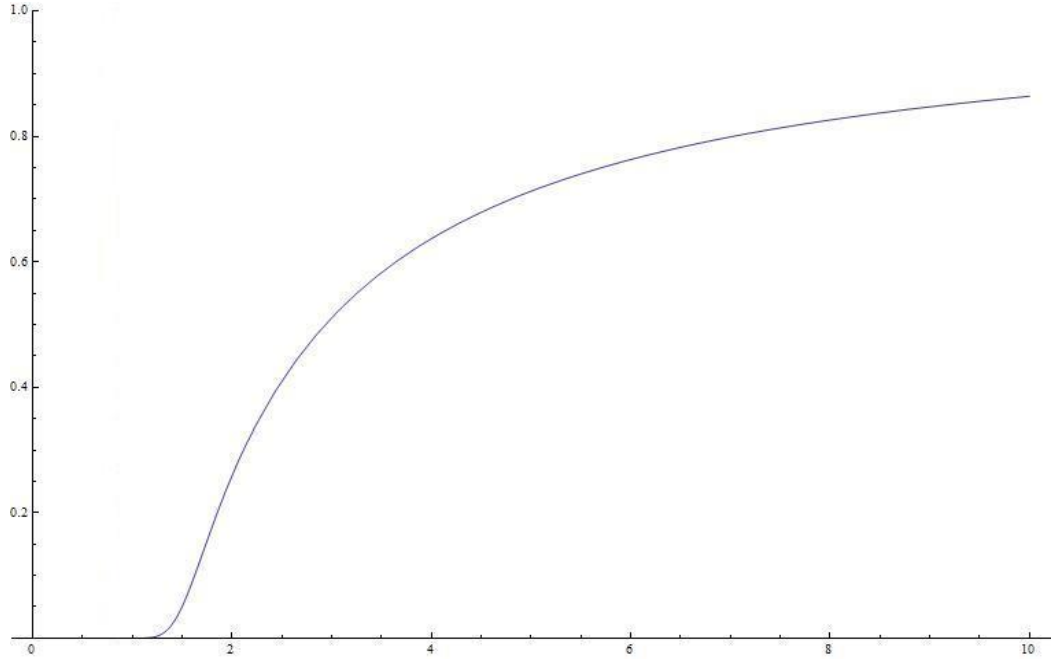

*Fig.5. Trans- isomer concentration distribution in the vicinity of beam focus. The abscissa axis represents coordinate in mkm, and ordinate axis represents relative concentration of trans- isomer.*

The characteristic induction time of this distribution is
$$t_{induction} = \frac{1}{B} \approx 10^{-8}(s),$$
And for its characteristic relaxation time we get:
$$t_{relax} \approx \frac{4R^2}{D} \approx 4 \cdot 10^{-2}(s)$$

**Processes in the cell culture surface treated with laser.**

The irradiated surface is described in materials and methods section 4. To estimate the trans- isomer concentration during the optical mapping the following diffusion model can be adopted: the cell culture of 5 mkm thickness is placed in the DMEM solution. Diffusion in the cells (e.g. parallel to the surface) is negligible compared to diffusion between cells and the solution above. Hence the concentration change done by the absorption at any point of the culture relaxates exponentially. According to our experiments this time is equal to several minutes, while the induction time is $10^{-2}$ s. The concentration change dependent on time can be described by the function
$$c = c_0 e^{\frac{-\sqrt{\pi}}{2} B\tau \cdot \left(erf\frac{t-t_0}{\tau} + erf\frac{t_0}{\tau}\right)} (5),$$
where $t_0$ is moment of maximum intensity in a pulse and $\tau$ is pulse duration.
This expression is illustrated by the graph on fig.6.

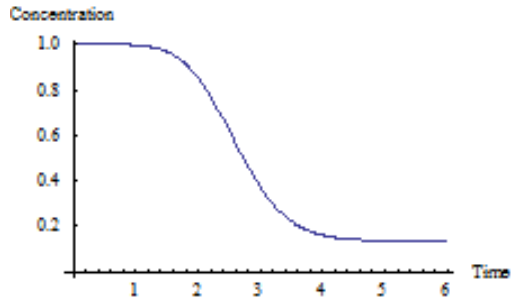

*Fig.6. Illustration of function (5).*

**Transformation of trans-AzoTAB by two-photon absorption inside excitable cardiac cells.**

The first stage of each experiment was devoted to control of cardiac cell culture quality. It included cardiomyocytes staining with fluo 4 calcium fluorescent dye and optical mapping of the culture using setup described in the Materials and methods section (see also fig. 1). If the culture demonstrated good confluency and homogenous excitability we used it in further. In the second phase of experiment we added 100 mkM AzoTAB in Tyrod's solution above cover slip with cardiomyocytes and conducted optical mapping procedure until excitation in all cells stopped. Usually it took about 1-5 minutes depending on the concrete cell culture. After that we chose the area for laser radiation and marked it. Duration of laser radiation procedure took from 30 to 105 minutes in accordance with the method described in previous section. After that the optical mapping procedure of the cardiac tissue exposed to laser had been conducted during the period of one minute. On the Fig. 7 the frame series of optical mapping video is shown. It demonstrates that after laser treatment only a small group of cells situated exactly in the exposed area have recovered excitability. The flashes of dye around this area are practically absent.

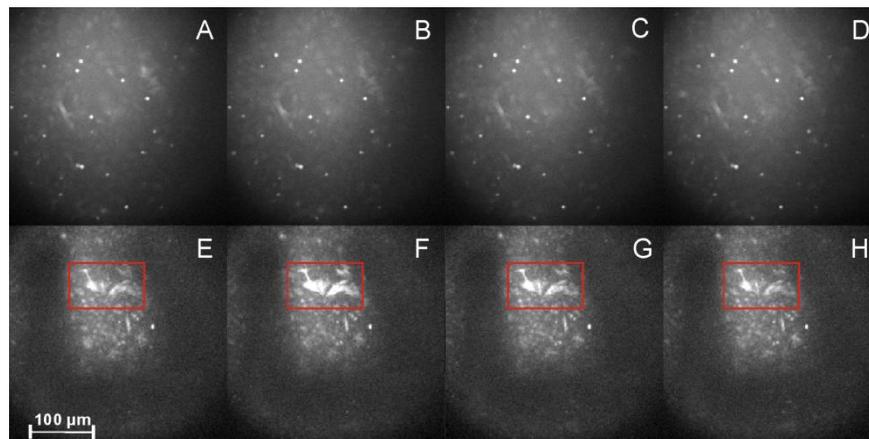

*Fig. 7.* Recovery of cardiac cells excitability after 30 minutes irradiation with IR-laser (red rectangle). The series A-D corresponds to the complete inhibition of cardiomyocytes auto excitability after incubation with 100 mkM AzoTAB. The series E-H was taken after 30 minutes exposition of culture cover slip to laser radiation. In the red rectangular the recovered contracting group of cardiomyocytes is marked. Concentration of ascorbinic acid in tyrod buffer was 0.2 mM.

The results of another experiment with prolonged laser exposition are presented on the fig. 8.

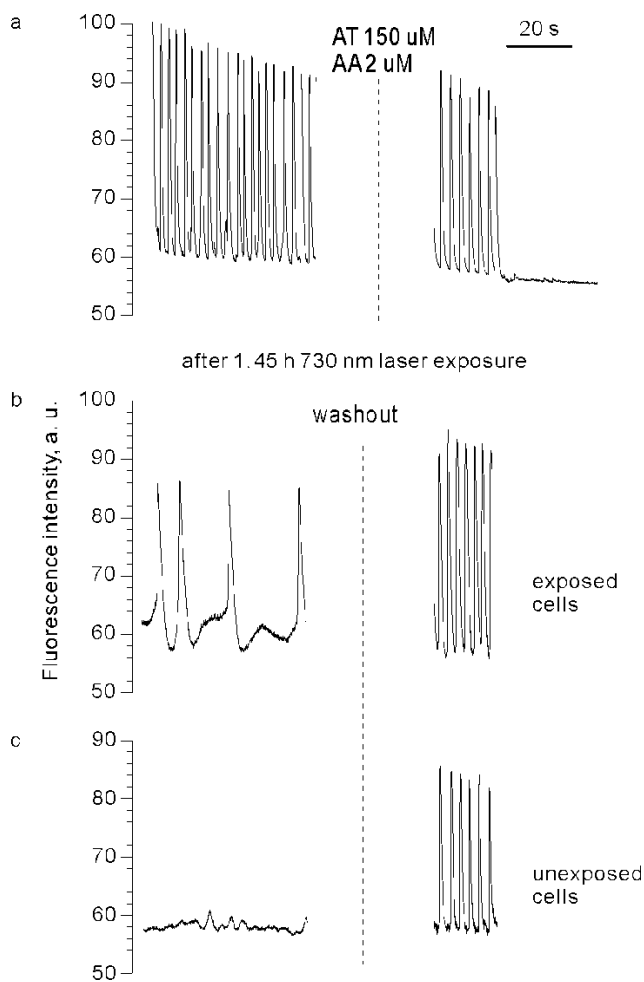

**Fig. 8.** Fluorescent emmision was measured from primary cardimyocytes in a cell network with spontaneous activity. When the tyrode salts buffer was replaced with the buffer containig 150 uM AzoTAB and 2 uM L-Ascorbic acid the activity ceased in a minute (a). Then the 100 um sized square region was exposed to 730 nm laser irradiation, and the activity of the exposed cells was partially restored (b) while the activity of the unexposed cells was not (c). After the washout the activity of all the cells was restored.

It demonstrates that recovery also takes place, but activity of the cells even in laser exposed area is rather low if compared with initial activity. We should mention here that in accordance with both experiments with cell cultures and theoretical accounts for bulk aqueous phase the prolonged laser treatment on trans-AzoTAB in darkness over 30 minutes doesn't give any sufficient increase of cis-AzoTAB concentration in the medium, in a 30 of exposition the system almost comes to its steady state.

**Conclusion.**
Finally we can summarise that two-photon absorption technique can be adopted for the optical control of cell excitability using azobenezene-based photosensitizers. It is important to denote that IR-radiation allows to restore selectively contractility of the cells placed deep inside myocardium without touching outer surface layers. From our viewpoint current results are important for the studies devoted to the structure-function organisation of complex three dimensional biological excitable systems.